\documentclass[manuscript]{aastex61}

\newcommand\aastex{AAS\TeX}

\usepackage{amsmath}

\received{}
\revised{}
\accepted{}
\shorttitle{\aastex\ sample article}
\shortauthors{Guo et al.}

\begin{document}

\title{Wave Heating in simulated multi-stranded coronal loops}

\correspondingauthor{Mingzhe Guo}
\email{mingzhe.guo@kuleuven.be}

\author{Mingzhe Guo}
\affiliation{Institute of space sciences, Shandong University, Weihai 264209, China}
\affiliation{Centre for mathematical Plasma Astrophysics, Department of Mathematics, KU Leuven, 3001 Leuven, Belgium}

\author{Tom Van Doorsselaere}
\affiliation{Centre for mathematical Plasma Astrophysics, Department of Mathematics, KU Leuven, 3001 Leuven, Belgium}

\author{Konstantinos Karampelas}
\affiliation{Centre for mathematical Plasma Astrophysics, Department of Mathematics, KU Leuven, 3001 Leuven, Belgium}

\author{Bo Li}
\affiliation{Institute of space sciences, Shandong University, Weihai 264209, China}

%\collaboration{(LaTeX collaboration)}

%% Note that the \and command from previous versions of AASTeX is now
%% depreciated in this version as it is no longer necessary. AASTeX 
%% automatically takes care of all commas and "and"s between authors names.

%% AASTeX 6.1 has the new \collaboration and \nocollaboration commands to
%% provide the collaboration status of a group of authors. These commands 
%% can be used either before or after the list of corresponding authors. The
%% argument for \collaboration is the collaboration identifier. Authors are
%% encouraged to surround collaboration identifiers with ()s. The 
%% \nocollaboration command takes no argument and exists to indicate that
%% the nearby authors are not part of surrounding collaborations.

%% Mark off the abstract in the ``abstract'' environment. 
\begin{abstract}

It has been found that the Kelvin-Helmholtz instability (KHI) induced by both transverse and torsional oscillations in coronal loops can reinforce the effects of wave heating.
In this study, 
we model a coronal loop as a system of individual strands,
and we study wave heating effects by considering a combined transverse and torsional driver at the loop footpoint.
We deposit the same energy into the multi-stranded loop and an equivalent monolithic loop,
and then observe a faster increase in the internal energy and temperature in the multi-stranded model.
Therefore,
the multi-stranded model is more efficient in starting the heating process.
Moreover,
higher temperature is observed near the footpoint in the multi-stranded loop and near the apex in the monolithic loop.
The apparent heating location in the multi-stranded loop agrees with the previous predictions and observations.
 Given the differences in the results from our multi-stranded loop and monolithic loop simulations, and given that coronal loops are suggested to be multi-stranded on both theoretical and observational grounds, our results suggest that the multi-strandedness of coronal loops needs to be incorporated in future wave-based heating mechanisms.

\end{abstract}


\section{INTRODUCTION} 
\label{sec_intro}

Various magnetic structures in solar atmosphere are reported to support a large amount of magnetohydrodynamic (MHD) waves and oscillations \citep[e.g.,][]{2012RSPTA.370.3193D,2016SSRv..200...75N}, 
which are believed to be an important candidate to heat the solar corona due to their capability of carrying energy \citep[e.g.,][]{2009SSRv..149..229T, 2012RSPTA.370.3217P}. 
To this end,
dissipation mechanisms are required to transfer the wave energy into internal energy. 
For kink oscillations,
resonant absorption is expected to transfer the collective transverse modes into local azimuthal modes in an inhomogeneous region \citep{1988JGR....93.5423H,1992SoPh..138..233G,2011SSRv..158..289G}.
Meanwhile,
phase mixing of Alfv{\'e}n modes between different magnetic surfaces help the energy cascade from large spatial scale structures to small structures,
then the wave energy dissipates at such small structures \citep{1983A&A...117..220H,2015ApJ...803...43S,2019ApJ...870...55G}.
Recently, 
numerical studies have found that transverse waves in coronal loops can induce the Kelvin-Helmholtz instability (KHI),
due to the strong velocity shear near the edge of the loops \citep[e.g.,][]{2008ApJ...687L.115T,2014ApJ...787L..22A,2015A&A...582A.117M, 2017A&A...604A.130K,2017A&A...602A..74H,2017A&A...607A..77H,2019ApJ...870...55G}. 
The generated small turbulent structures can help dissipate wave energy more easily.
Therefore,
Ohmic or viscous dissipation of wave energy can be achieved \citep{1990ApJ...360..279P,1991PhRvL..66.2871P, 1995ApJ...444..471O} thanks to the occurrence of small structures in the aforementioned physical processes.

From the early observations with the Transition Region and Coronal Explorer (TRACE) to the recent measurements with high resolution instruments (e.g., SDO/AIA, Hi-C), 
more and more evidence shows that coronal loops are filled with bundles of thin strands \citep{2002ApJ...580.1159T, 2012ApJ...755L..33B, 2013A&A...556A.104P, 2013Natur.493..501C,2017ApJ...840....4A}.
Multi-stranded loop models can help to explain some observations,
which thus provide an indirect evidence for the loop fine structuring.
\citet{2010ApJ...719..576G} modelled the loops composed of subarcsecond strands to explain the ``fuzzy" appearance of hot loops \citep{2009ApJ...694.1256T}.
Meanwhile,
they predicted that the fuzziness will decrease in strands with temperature larger than 3MK,
which has been confirmed with SDO/AIA observations in active region by \citet{2011ApJ...736L..16R}.
Transverse waves are also reported in the thin threads observed with Hinode Solar
Optical Telescope (SOT) \citep{2008A&A...482L...9O}. 
 Analytically,
 transverse oscillations in a two loop system have been investigated by \citet{2008ApJ...676..717L, 2008A&A...485..849V}.
Oscillations and flows in a more complicated twisted multithreaded model has been studied by \citet{2009ApJ...694..502O}.
 \citet{2010ApJ...716.1371L} investigated the transverse oscillations in a loop system with randomly distributed strands and found that the interactions of strands strongly influence the dynamics of the whole loop and thus disturb the coherent motions of the strands.
 In such randomly structured loops,
 mode coupling between kink modes and Alfv{\'e}n modes is still efficient \citep{2008ApJ...679.1611T,2011ApJ...731...73P}. 
 
 Simulations of a tightly packed multi-stranded loop by \cite{2016ApJ...823...82M} have shown that 
 the loop is unstable when driving by transverse waves.
 \citet{2018A&A...610L...9K} revealed that a driven monolithic loop can become fully deformed due to the induced instability.
 Therefore,
 the plasma in both multi-stranded and monolithic models can achieve a turbulent state with continuous driving.

In this paper,
we will examine the wave heating effects from driven transverse and Alfv{\'e}n oscillations in a multi-stranded loop.
Meanwhile,
a density equivalent monolithic loop is also considered,
in order to find out how the loop configuration will influence the heating effects.

\section{NUMERICAL MODELS}
\label{sec_models}

\subsection{Equilibrium and Drivers}
\label{sec_sub_ini}

\begin{figure*}
\gridline{\fig{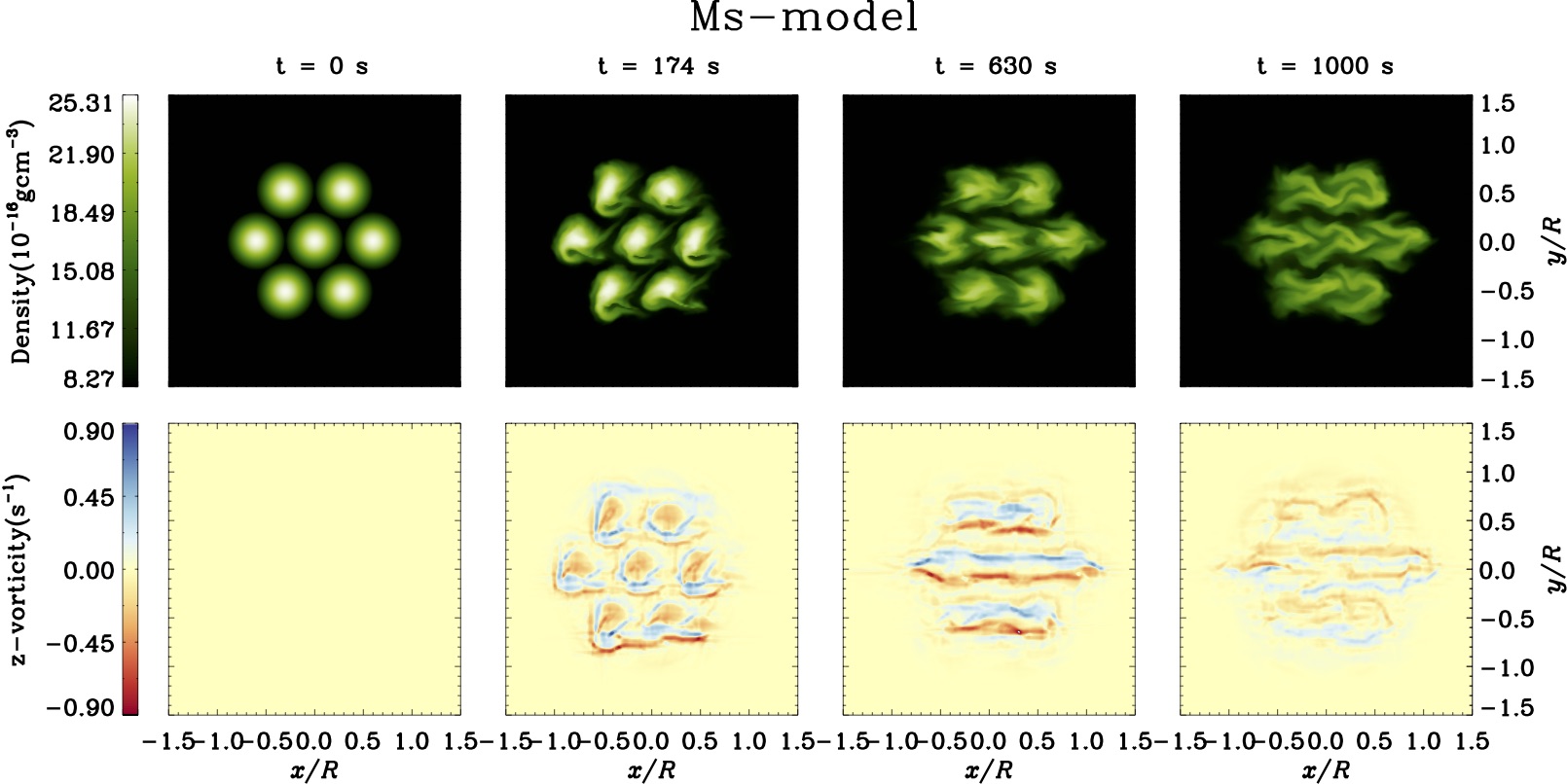}{0.8\textwidth}{(a)}}
\gridline{\fig{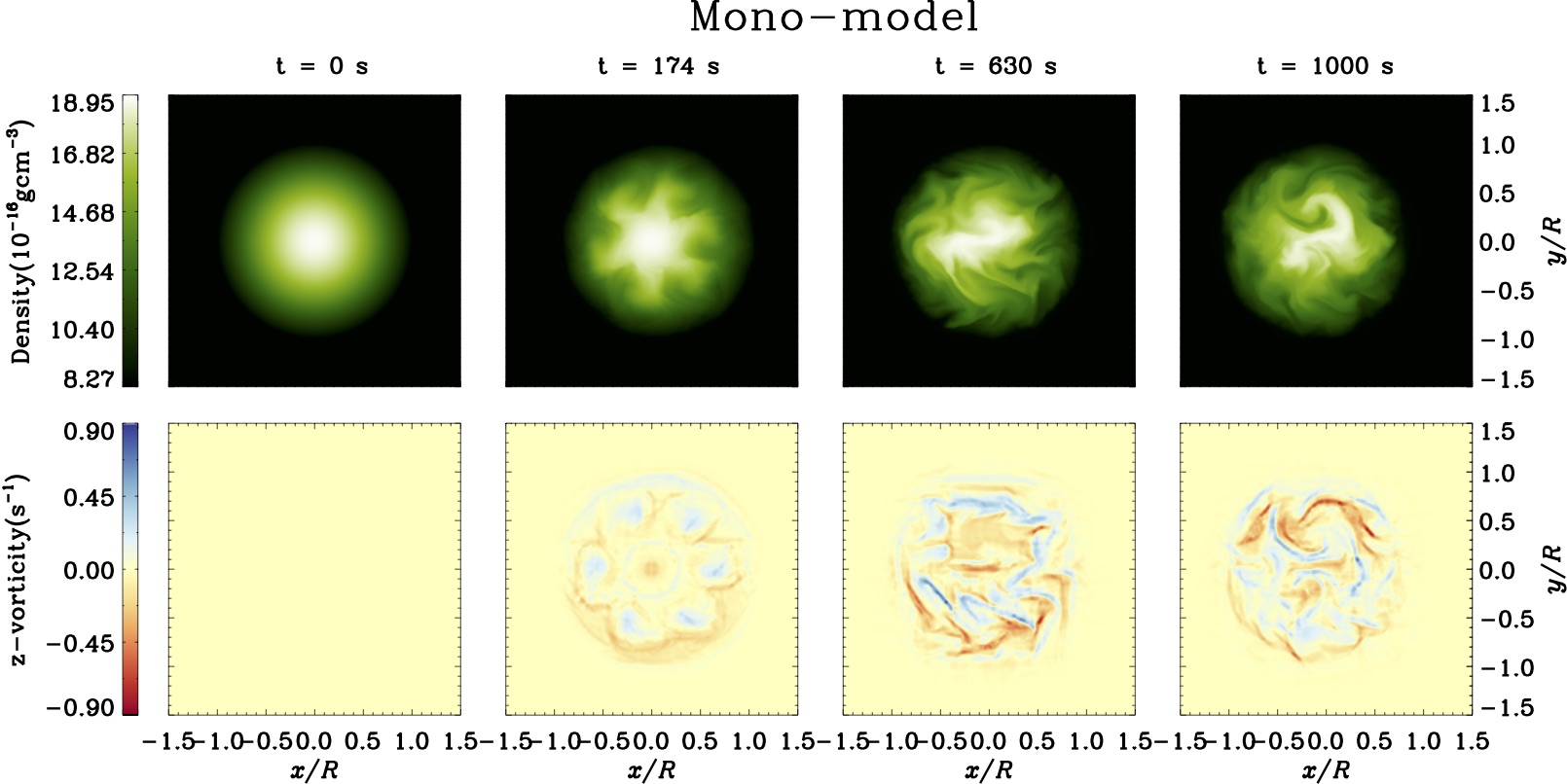}{0.8\textwidth}{(b)}}
\caption{Snapshots of density (upper row) and $z$-vorticity (lower row) evolutions of the cross-section at loop apex for the Ms-model (a) and Mono-model (b). 
}					
\label{fig1}	
\end{figure*}

In our simulations, 
we consider a loop system with density enhanced,
straight strands,
which are tightly packed and embedded in a uniform background corona.
A uniform temperature $T=1{\rm MK}$ is considered in the whole simulation domain.
To maintain magnetostatic pressure balance, 
the magnetic field has a slight variation from the center of each strand $B_{\rm 0} = 50 {\rm G}$ to the external medium $B_{\rm e} = 50.07 {\rm G}$.
For simplicity,
the loop is filled with seven identical strands.
The initial loop cross-section can be found in the left panel of Figure \ref{fig1}.
Each strand has a radius $R_{\rm s}=0.3{\rm Mm}$ and an initial peak density $\rho_{\rm p}=3\rho_{\rm e}$,
where the external background density is $\rho_{\rm e}=8.36\times10^{-16} {\rm g~cm^{-3}}$.
The density profile in each strand is given by
\begin{equation}
\rho_{\rm s}(r_n)=\rho_{\rm e}+(\rho_{\rm p}-\rho_{\rm e})\cos\left(\frac{\pi r_n}{2R_{\rm s}}\right),
r_n\le R_{\rm s},
\label{eq_rhos}
\end{equation}
where $r_n=\sqrt{(x-x_n)^2+(y-y_n)^2}$, $x_n,y_n$ is the center location of each strand, $n=1,2,...,7$.
The loop length is $L=150{\rm Mm}$, which is chosen within the range of observations \citep{2002SoPh..206...99A}.
As a comparison,
we also consider a density equivalent monolithic loop by redistributing the plasma according to Equation \ref{eq_rhos} but in a radius $R=1{\rm Mm}$ region.
Then the ratio between the peak and the background density can be obtained by
\begin{equation}
\alpha=\displaystyle\frac{\pi^2}{4\pi-8}\left\{\displaystyle\frac{1}{\pi R^2\rho_{\rm e}}\left[\displaystyle\sum\limits^7_{n=0}\int\int_0^{R_{\rm s}}\rho_{\rm s}(r_n)r_ndr_nd\phi+\rho_{\rm e}\left(\pi R^2-7\pi R_{\rm s}^2\right)\right]-1\right\}+1,
\label{eq_rho_mono}
\end{equation}
which gives $ \alpha=2.26$.
 Thus the peak density of the monolithic loop is $\rho_{\rm p}'= 2.26\rho_{\rm e}$.
 The terms in the square brackets in Equation \ref{eq_rho_mono} are the integration of density of the seven strands and the background medium density between them in the radius $R=1{\rm Mm}$ region in the multi-stranded model.
The other terms come from the integration of density in the same region in the monolithic loop.

For both models, 
we employ a mixed driver that consists of a loop-region transverse motion and seven independent strand-region torsional motions at the footpoint ($z=0$).
The transverse motion is similar to those in the previous models \citep{2010ApJ...711..990P, 2017A&A...604A.130K, 2019ApJ...870...55G},
which is a continuous, monoperiodic ``dipole-like" driver and given by
\begin{equation}
\mathbf{v}=\mathbf{v}_{\rm e}+\displaystyle \frac{1}{2} \left\{1-\tanh \left[b(\sqrt{x^2+y^2}/R-1)\right] \right\}(\mathbf{v}_{\rm i}-\mathbf{v}_{\rm e}),
\label{eq_v}
\end{equation}
where 
\begin{equation}
\mathbf{v}_{\rm i} =v_0\left[\sin\left(\frac{2\pi t}{P_{\rm k}}\right),0,0\right],
\label{eq_vi}
\end{equation}
\begin{equation}
\mathbf{v}_{\rm e} =v_0 R^2 \sin\left(\frac{2\pi t}{P_{\rm k}}\right)\left[ \frac{x^2-y^2}{(x^2+y^2)^2},\frac{2xy}{(x^2+y^2)^2},0 \right],
\label{eq_ve}
\end{equation}
where $b=16$,
which gives the width of the intermediate layer $l\approx0.2{\rm Mm}$.
$v_0$ is the velocity amplitude,
which varies in different models.
The period is $P_{\rm k}=87{\rm s}$ ($\textbf{78.5}{\rm s}$) for the multi-stranded model (the monolithic model),
which corresponds to the analytical value for the fundamental kink mode \citep{1983SoPh...88..179E}.
For the multi-stranded model,
an initially perturbed loop test shows that the kink periods of the oscillations of individual strands are almost the same; since the strands here are tightly packed.
Simultaneously,  
the torsional motions at the footpoint ($z=0$) are described by 
\begin{equation}
v_{\theta n}=v_0 \sin\left[\frac{2\pi t}{P_{\rm A}(r_n)}\right], r_n\le R_{\rm s},
\label{eq_vtheta}
\end{equation}
where $n=1,2,...,7$.
The period is given by
$P_{\rm A}(r_n)=2L\sqrt{\mu_0 \rho(r_n)}/B(r_n)$.
 In both models, 
the drivers follow the motions of loops. 
The positions change with time,
making sure that the motions at footpoint are always described by Equation \ref{eq_v} and Equation \ref{eq_vtheta}. 

\subsection{Numerical setup}
\label{sec_sub_setup}

To solve the 3-D ideal MHD equations, 
we use the PLUTO code \citep{2007ApJS..170..228M},
in which a second-order finite volume piecewise parabolic method (PPM) is employed for the spatial integration \citep{2014JCoPh.270..784M}.
The numerical fluxes are computed by a Roe Riemann solver.
A third-order Runge-Kutta algorithm is used for the time advance. 
The simulation domain is ${\rm \left[-8,8\right] Mm \times \left[-8,8\right] Mm \times \left[0,150\right] Mm}$.
We adopt a uniform grid of 100 points from $0$ to $L$ in the $z$-direction and 256 non-uniformly spaced cells in the $x$- and $y$-directions.  
The highest resolution is ${\rm 15~km}$ in the region of $|x,y| \leq 1.5 {\rm Mm}$.

We fix the velocities at $z=L$ to be zero to mimic loops anchored in the lower atmosphere. 
At the other footpoint ($z=0$),
the $z$-component velocities are antisymmetric and $v_x,v_y$ are described by the driver.
The other variables at both footpoints are set to be Neumann-type (zero-gradient) conditions.
All the lateral boundaries are set to be outflow conditions. 
  
\section{RESULTS}
\label{sec_results}

We ran our simulations until $t=1000{\rm s}$ for both the multi-stranded model (``Ms-model'' hereafter) and the monolithic model (``Mono-model'' hereafter). 
In the following analysis,
we focus on the subvolume of $|x,y|\leq 1.5{\rm Mm},0\leq z\leq 150{\rm Mm}$.

In the previous study,
we found that the energy input into the system was influenced by the perturbations of the magnetic field at footpoint \citep{2019ApJ...870...55G}.
This means that the input energy varies in different models even if the same velocity drivers are employed.  
For a straightforward comparison,
in the current study,
we let the input energy in both models to be the same through keeping the velocity amplitude of the driver $v_0=4{\rm km~s^{-1}}$ in the Ms-model,
while $v_0=2.8{\rm km~s^{-1}}$ in the Mono-model by a parameter study.
Therefore, 
any variations of the internal energy and temperature are not induced by a difference in the input energy.

The evolution of the loop cross-section at the apex is shown in Figure \ref{fig1}.
It indicates that the Kelvin-Helmholtz instability (KHI) is induced in both models.
In the Ms-model,
the KHI is quickly induced by the mixed transverse and torsional motions at each strand,
and the intermixing between different strands.
In the Mono-model,
besides the collective transverse motion,
Alfv{\'e}n modes are also established inside the loop,
which help the instability extend from the loop edge to almost the whole loop region.

\begin{figure*}
\gridline{\fig{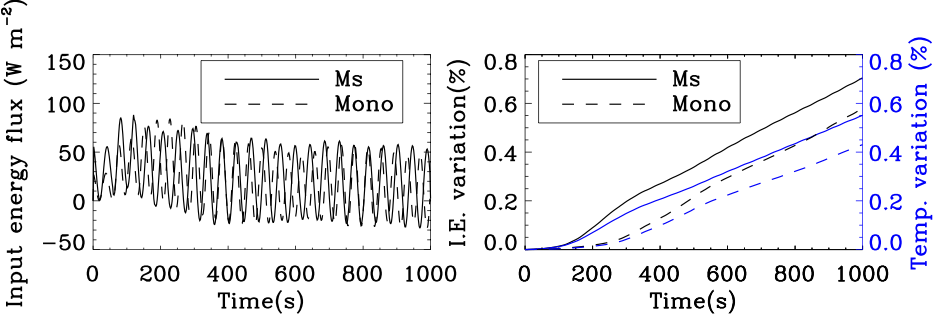}{0.9\textwidth}{}}
\caption{Left: the input energy flux variations. 
Right: percentages of volume-averaged internal energy (black) and temperature (blue) variations.}
\label{fig2}	
\end{figure*}

To quantify the heating effects from wave energy dissipation,
we examine the internal energy and temperature variation in both models.
The input energy flux from the driver is given by
\begin{equation}
F(t)=-\frac{1}{A}\int_{A} \textbf{S}\cdotp d\textbf{A},
\label{eq_energyflux}
\end{equation}
where $\textbf{S}=\textbf{E}\times\textbf{B}/\mu_0$ is the Poynting flux, 
$A$ is the surface area of the subregion ($|x,y|\leq 1.5{\rm Mm}$) and $d\textbf{A}$ is the normal surface vector of the bottom plane.
As mentioned before,
the input energy flux in both models are at the same level.
From the left panel of Figure \ref{fig2},
we see that the energy flux is about $50 {\rm W m^{-2}}$ in the beginning and $\sim20 {\rm W m^{-2}}$ near the end of the simulation. 
It is somewhat smaller than the radiative energy losses of the quiet corona, $\sim100 {\rm W m^{-2}}$ \citep{1977ARA&A..15..363W, 2007Sci...317.1192T}.
Further discussions can be found in Section \ref{sec_conclusion}.
From the right panel of Figure \ref{fig2},
 we can see that the volume-averaged internal energy and temperature increase in both models.
 In the Ms-model,
 both quantities have a faster increase before $300{\rm s}$ and then keep a stable increase rate in the later stage.
 The increase rate in the Mono-model has a slight variation and it is almost the same as that in the Ms-model after $t=550{\rm s}$.
 This indicates that both models are effective in heating,
 however,
 the multi-stranded loop is more efficient in starting the heating process than the equivalent monolithic loop.
 The kink frequency in the two models is different due to the different peak density,
  leading to the phase difference in the left panel of Figure \ref {fig2}.
  However,
  such a frequency difference is quite small,
  we thus have 11.5 periods kink oscillations in the multi-stranded model and 12.8 periods kink oscillations in the monolithic model in the entire simulation time (1000{\rm s}).
  
\begin{figure*}
\centering
\gridline{\fig{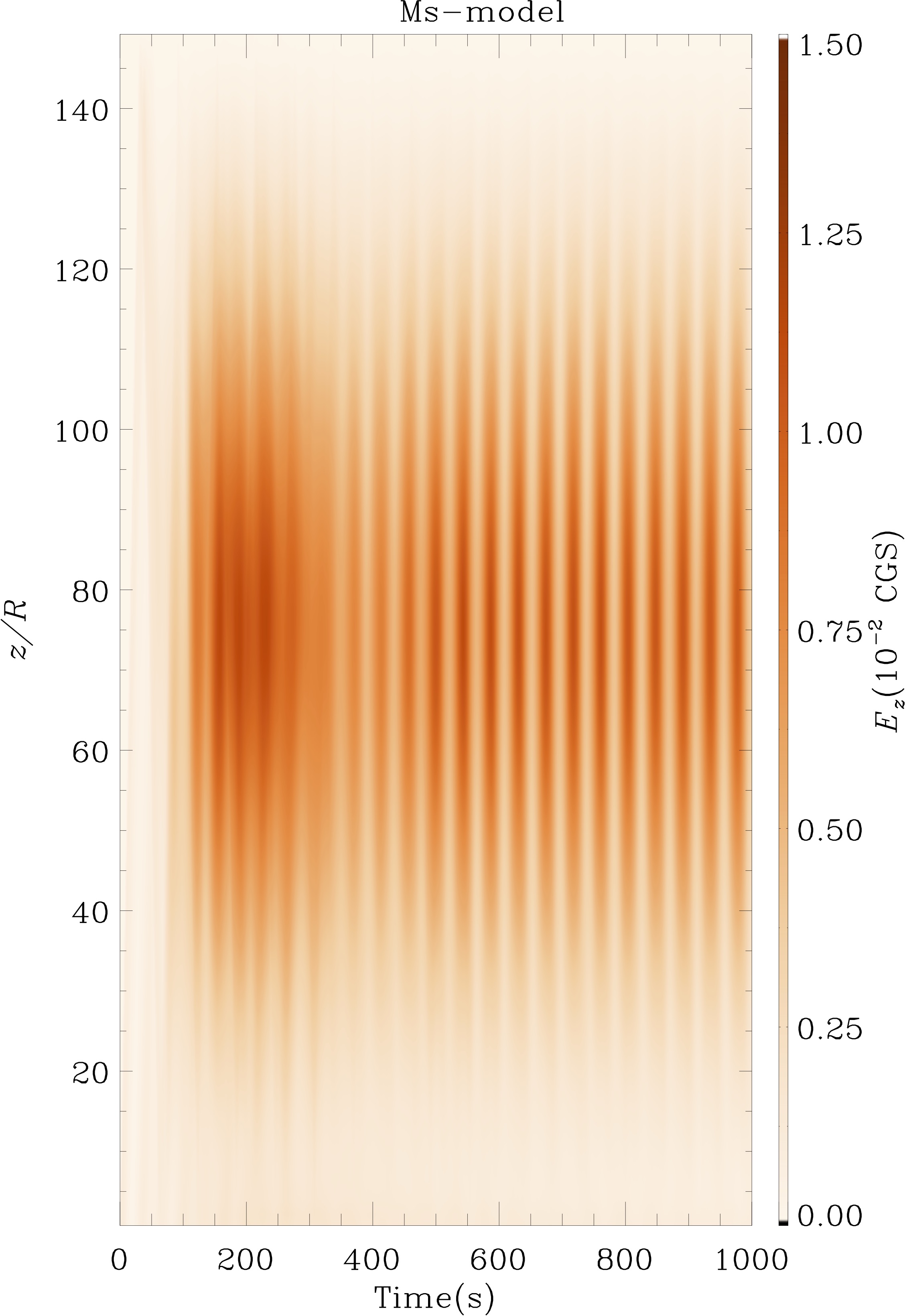}{0.33\textwidth}{}
              \fig{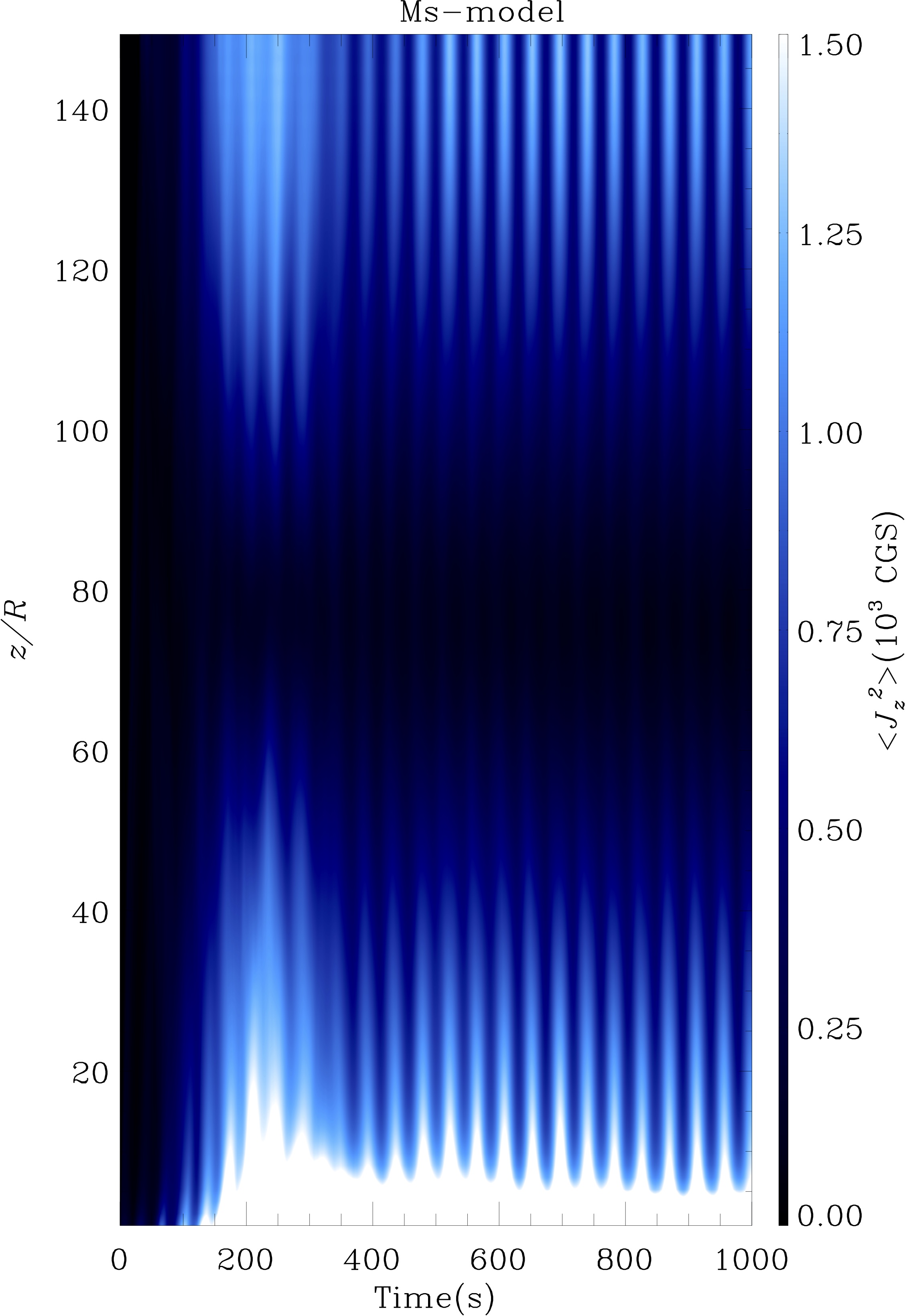}{0.33\textwidth}{}
              \fig{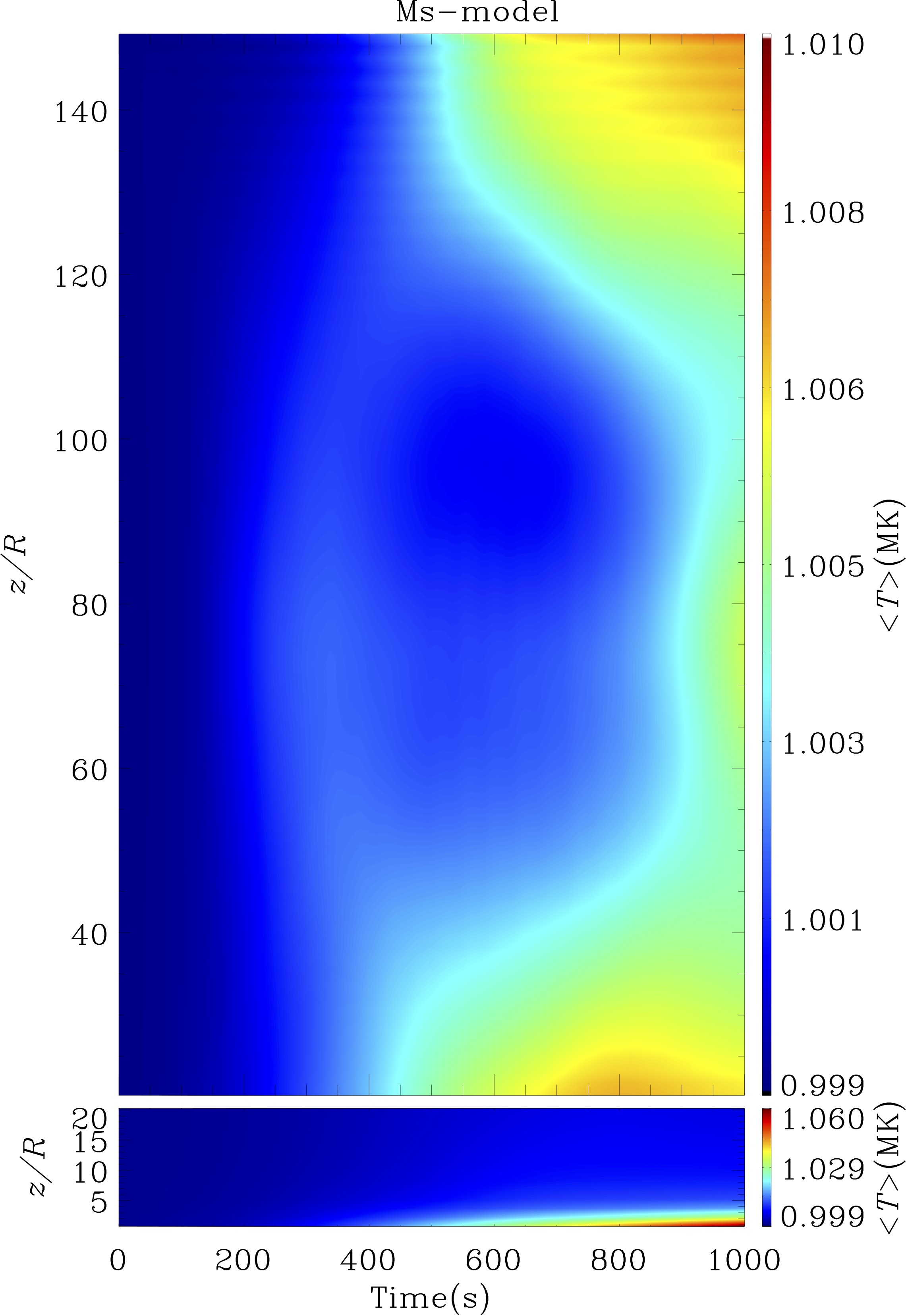}{0.33\textwidth}{}
           }
\gridline{\fig{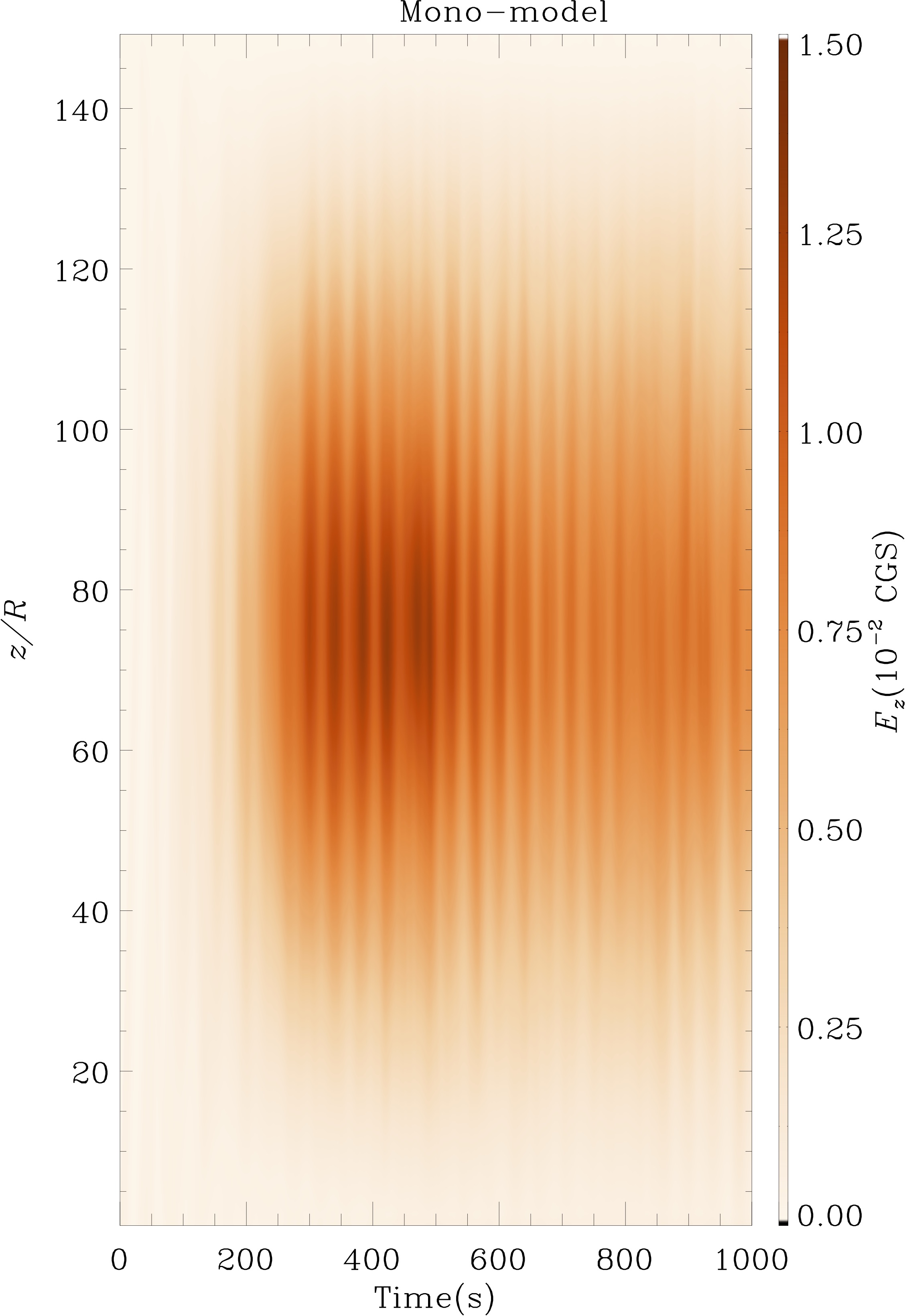}{0.33\textwidth}{}
              \fig{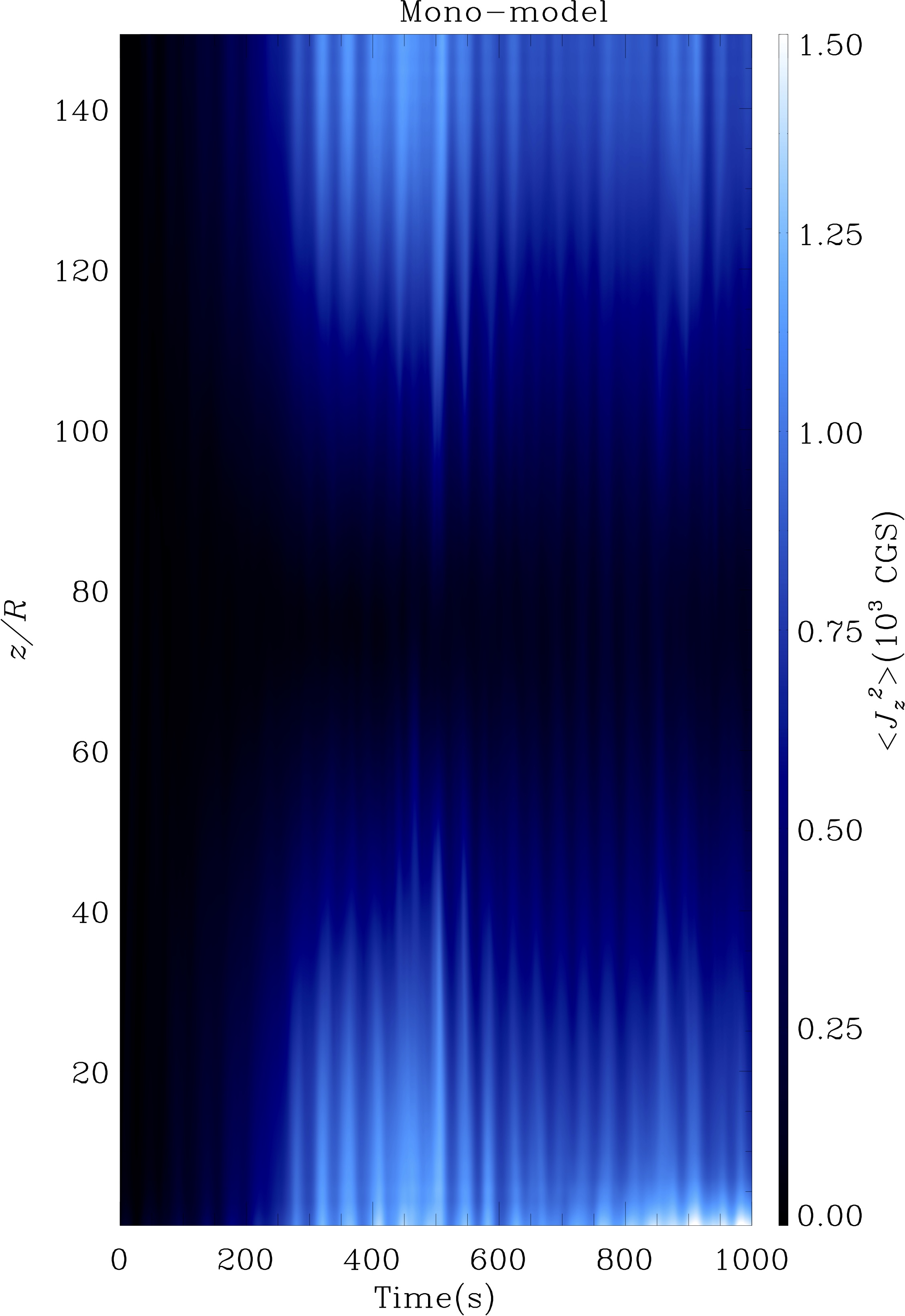}{0.33\textwidth}{}
              \fig{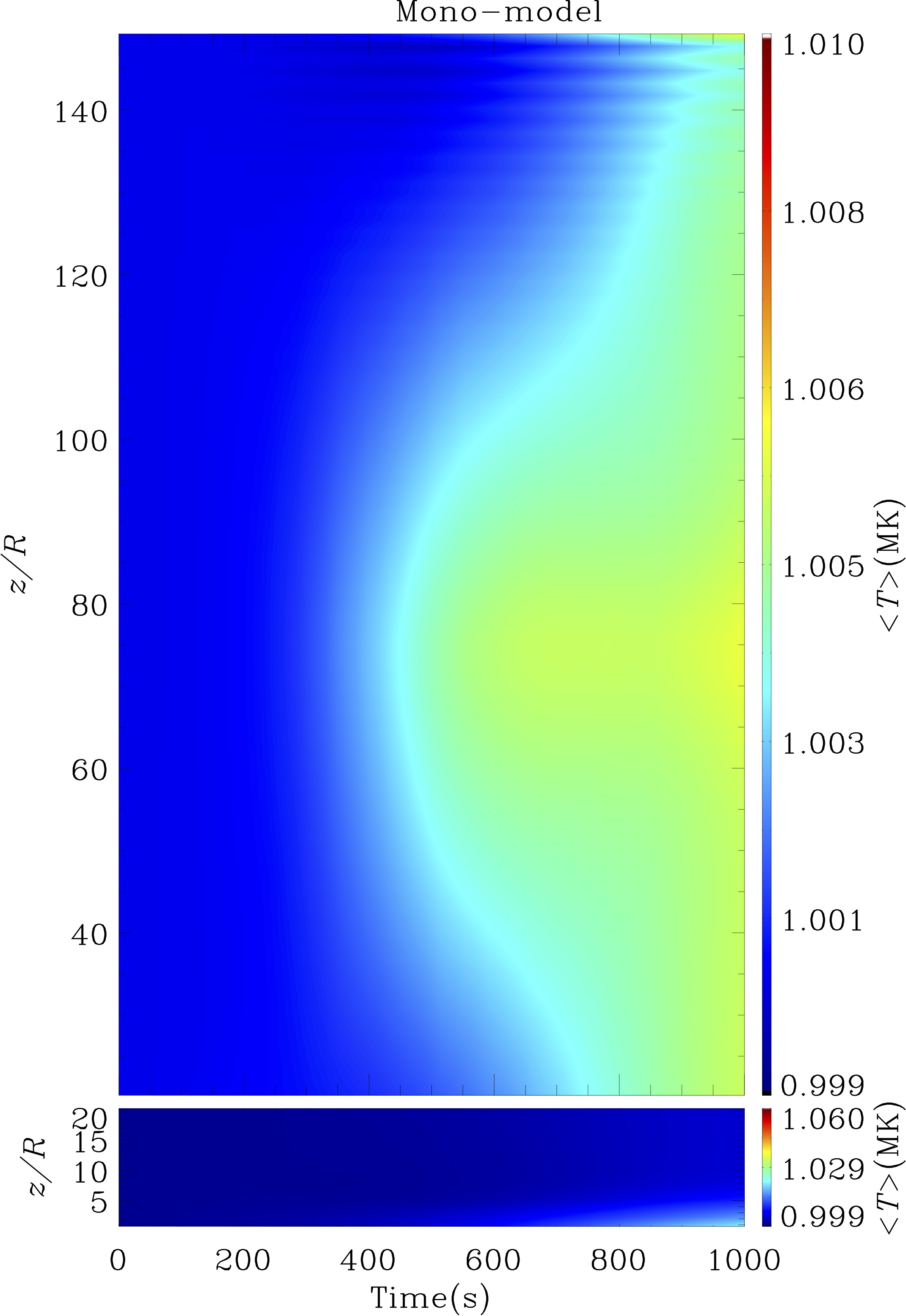}{0.33\textwidth}{}
           }              
\caption{Averaged enstrophy ($E_z$), surface averaged square $z$-current density ($J^2_z$), and surface averaged temperature profiles along the $z$-direction for the Ms-model and the Mono-model.}	
\label{fig3}				
\end{figure*}

 Now we go further to examine the heating properties in detail in both models, 
 rather than volume averaged values.
 In order to clarify the dissipation mechanisms in our models,
 we examine the averaged enstrophy ($E_z=\frac{1}{A}\int_{A} (\nabla \times v)_z^2 dA$) and the averaged square $z$-current density ($J^2_z$) profiles along the $z$-direction,
 which are shown in Figure \ref{fig3}.
As predicted in \citet{2007A&A...471..311V},
 we can see the maximum $E_z$ near the loop apex and the maximum $J^2_z$ near the loop footpoint in both models.
 The profiles do not exactly follow a sinusoidal profile and this is probably due to the driving of the plasma and the non-linear effects in our models.
 If the heating location is near the footpoint,
 then the heating is mainly caused by the high current density there,
 and thus the dominant heating mechanism there is resistive heating.
 Otherwise,
 if the heating location is near the loop apex,
 the dominant heating mechanism is viscous heating.
 In the Ms-model,
 the temperature is higher near the loop footpoint,
 which probably means that the resistive heating is dominant.
 This result seems to favor the the footpoint heating prediction and observations in \citet{2007A&A...471..311V}.
 We see that $J^2_z$ near the footpoint is larger in the Ms-model than that in the Mono-model.
 Due to the more turbulent structure in the Ms-model,
the magnetic field is greatly disturbed and a larger radial variation is induced.  
In Figure \ref {fig3},
 it seems that the hot plasma accumulates near the footpoint in the Ms-model.
 To understand this point,
 we examine the enstrophy at $z=130{\rm Mm}$ in Figure \ref{fig5}.
 As mentioned in \citet{2019ApJ...870...55G},
 the development of the turbulent structures induce a reduction in the vorticity evolution profile,
 because of the expansion of the smaller structures.
 A larger reduction indicates that smaller and smaller structures expand to a larger region and the plasma is more turbulent.
 In Figure \ref{fig5},
 we see that the turbulent structures are fully developed after about $t=300{\rm s}$ in the Ms-model and $t=500{\rm s}$ in the Mono-model.
The smaller turbulent structures develop faster in the Ms-model,
the heating process thus starts faster,
as is shown in Figure \ref{fig2}.
  The enstrophy in the Ms-model has a larger decrease than that in the Mono-model,
meaning that plasma in the Ms-model becomes more turbulent.
In the Ms-model,
we see that the maximum $J^2_z$ appears around $t=200{\rm s}$ in Figure \ref{fig3}.
It extends further along the $z$-direction from the footpoints towards the apex,
leading to a heated band from the footpoint to the apex.
Since the turbulent structures are not well developed around $t=200{\rm s}$,
it is thus just a slight temperature increase in the heated band.

\begin{figure*}
\centering
\gridline{\fig{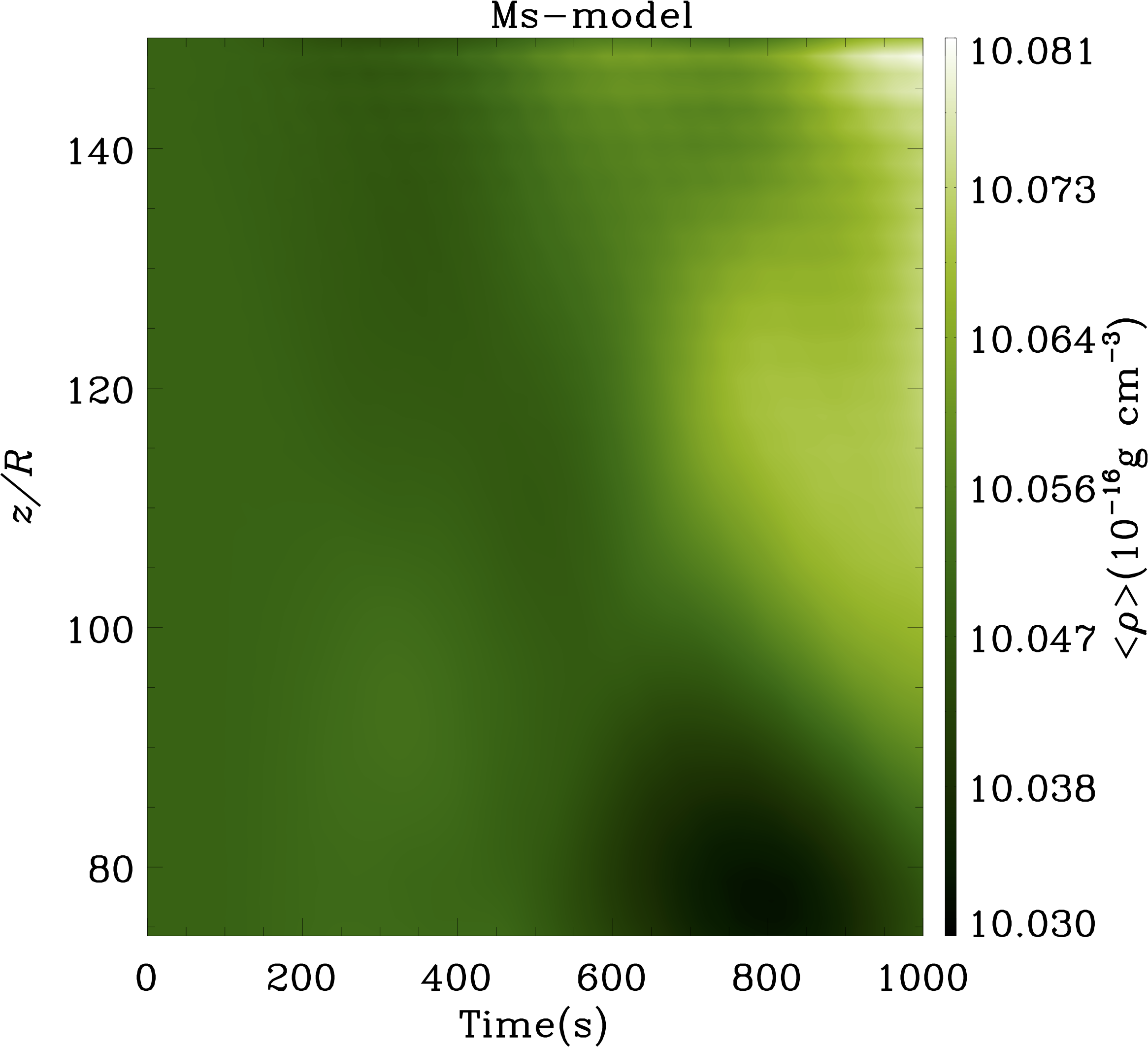}{0.49\textwidth}{}
              \fig{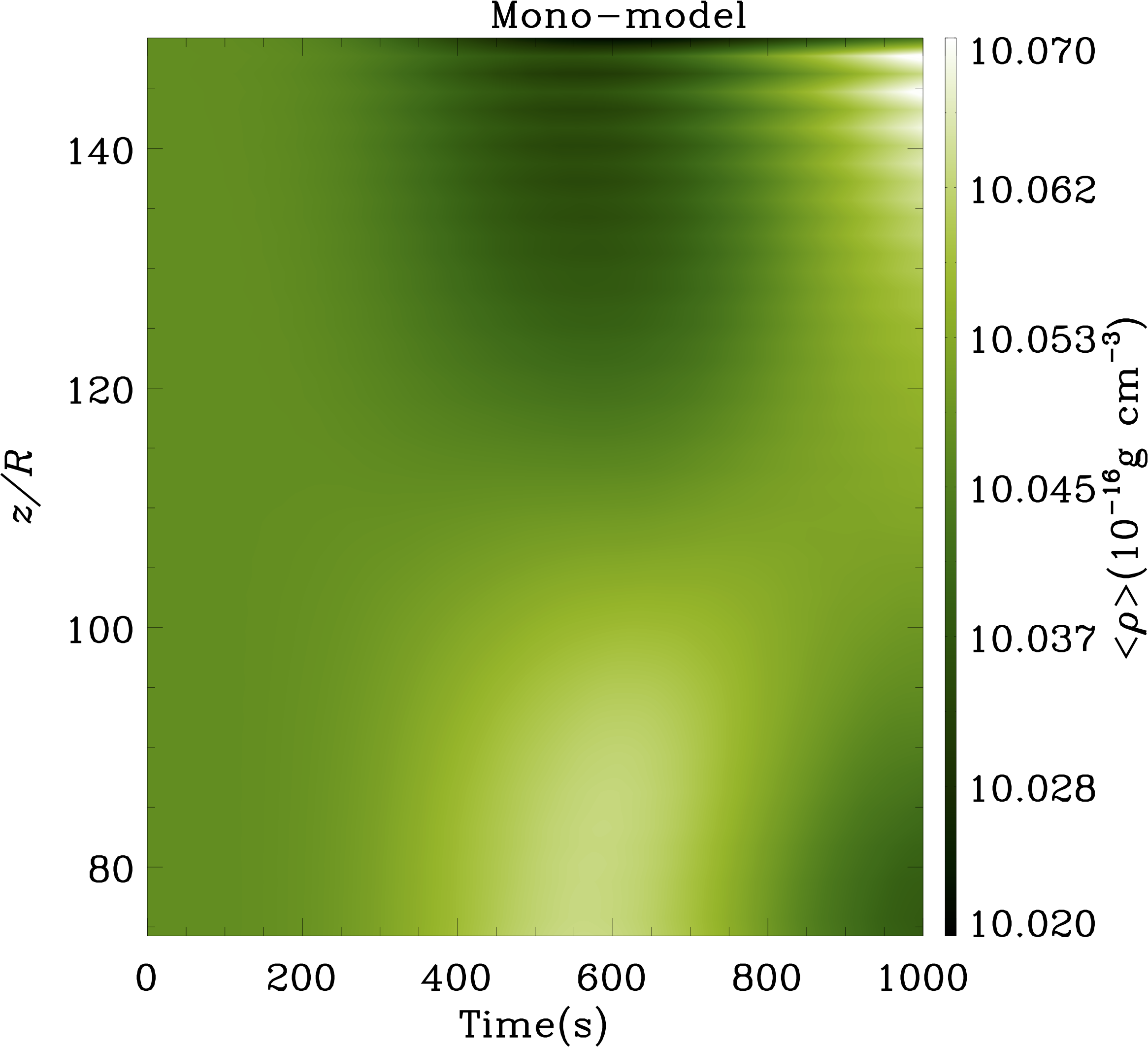}{0.49\textwidth}{}
           }              
\caption{Surface averaged density profiles along the $z$-direction from the apex ($z=75{\rm Mm}$) to the footpoint ($z=150{\rm Mm}$) for the Ms-model and the Mono-model.}	
\label{fig4}				
\end{figure*}

\begin{figure*}
\gridline{\fig{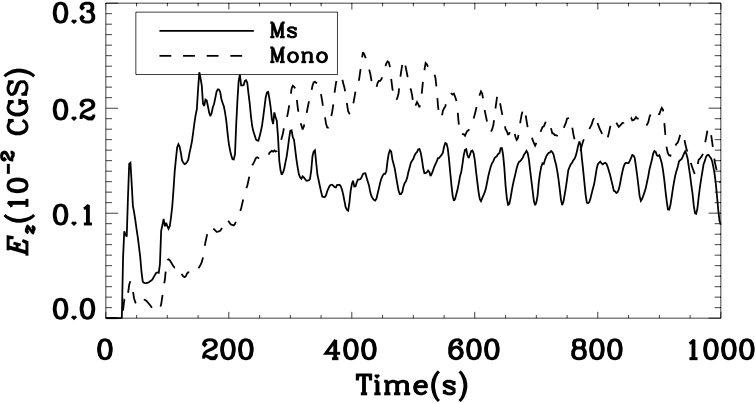}{0.6\textwidth}{}}
\caption{Time evolution of the enstrophy for the Ms-model (solid line) and Mono-model (dashed line). 
The quantities are averaged over the region of $|x,y|\leq 1.5{\rm Mm}$ at $z=130{\rm Mm}$. }					
\label{fig5}	
\end{figure*}

In the Mono-model,
the temperature is higher near the loop apex.
It is not only due to the higher vorticity at the apex than that in the Ms-model.
We can also observe an additional density fluctuation in the Mono-model in Figure \ref{fig4}.
Such periodic fluctuation has also been found in \citet{2016A&A...595A..81M, 2019A&A...623A..53K},
which is associated with the ponderomotive force in the case of standing oscillations in a loop \citep{2004ApJ...610..523T}.
This fluctuation can also influence the temperature profile in the Mono-model in Figure \ref {fig3},
inducing a temperature variation around $t=600{\rm s}$.
However,
due to the more turbulent plasma in the Ms-model,
such ponderomotive force associated fluctuation is greatly prevented by the redistribution of the magnetic field.

\citet{2019A&A...623A..53K} observed a similar temperature profile near the footpoint and the apex in a transverse driven monolithic model.
In the current work,
we obtain a larger heating region which extends from the footpoint to the apex in the Mono-model.
This is due to the inclusion of torsional motions in the current model.
The induced KHI extends the non-uniform layers where phase mixing takes place.
In \citet{2019ApJ...870...55G}, 
we proved that the mixed transverse and torsional motions induce more turbulent structures,
which can help dissipate wave energy of the excited kink and Alfv{\'e}n modes.
The energy dissipation is thus enhanced,
comparing to the single wave mode driving experiments.

In Figure \ref {fig4}, 
we notice that the averaged density at the apex decreases after about $t=600{\rm s}$ in the Ms-model
and $t=800{\rm s}$ in the Mono-model.
 This is similar to the enstrophy reduction in Figure \ref{fig5}.
 The extension of the smaller and smaller structures at a given height leads to a decrease in the surface averaged value,
especially near the loop apex.

 In Figure \ref{fig3},
 the temperature profiles near the lower footpoint ($z=0$) in both models are plotted separately due to their higher values when getting very close to the driver ($z< 5{\rm Mm}$).
 This is because the velocity shear between the different torsional driving regions induces extremely high current density,
 which can not be seen near the other fixed footpoint.
 Although this boundary effect slightly increases the temperature in the lower half part of the loop,
 it does not influence the main properties of the heating profile mentioned above.

\section{DISCUSSION AND CONCLUSIONS}
\label{sec_conclusion}
 
 In this study, 
 we simulated a multi-stranded loop with a mixed footpoint driver,
 considering both transverse and torsional motions.
 Through comparing with an equivalent monolithic loop,
we found that heating effects are observed in both models.
 The multi-stranded loop has a quick increase in the internal energy and temperature.
 Therefore,
 it is more efficient in starting the heating process than the monolithic model.
 Further studies showed that the main heating location is near the footpoint in the multi-stranded loop,
 while the temperature is higher near the loop apex in the monolithic loop.
 Therefore,
  the apparent heating location in the multi-stranded loop agrees with the footpoint heating prediction in the linear theory and observations.
  Considering the efficient heating effects,
  the multi-stranded loop probably can be a better choice to model coronal loops and study AC heating effects.
  
In this paper, 
we assume that both the monolithic and multi-stranded loops existed before the driving. 
So we just focus on the process when the drivers are launched. 
Perhaps it is true that the emergence time of different loops are different. 
However,
it seems more reasonable to consider a multi-stranded loop since the loop structure is filled with plasma according to the distribution of magnetic field due to the very low plasma beta. 
This means that the plasma moves together with the magnetic field. 
The ideal case is that once the loop structure is formed, 
the internal configuration of the loop is settled. 
So we do not need to consider a different construction time in the two models. 
However, 
since realistic coronal loops are highly dynamic, 
it would be not easy to recognise the construction time of the strands for the modern instruments. 
The observed fine structures in a loop can either be explained as waves induced instability (e.g., transverse wave induced KHI rolls suggested by \cite{2014ApJ...787L..22A}) or as sub-loops (strands).

Different velocity amplitude of drivers are employed in our models,
since the magnetic field at the footpoint evolves with time freely and it thus depends on the dynamics of the loop. 
Therefore, 
the Poynting flux depends on both the velocity described by the driver and the dynamics of the loop.
Similar to the enstrophy profile of Figure \ref{fig5},
a larger number of turbulent structures lead to a larger decrease in the averaged magnetic field perturbations at footpoint in the Ms-model due to the expansion of the turbulent structures.
Therefore,
to obtain an equivalent input energy as in the Mono-model,
a larger velocity in needed.
If we impose the same velocity, 
namely a larger velocity in the Mono-model,
the Poynting flux will probably increase. 
As a consequence, 
the values of internal energy or temperature in the later simulation time will be affected. 
However, 
if we focus on the start stage of the heating process,
we can hardly say that the monolithic loop would be easier to be heated. 
A faster start of the heating process (as in the Ms-model) depends on the faster development of the turbulent structures.
However,
the initial number of the KHI eddies will decrease when the shear velocity between a loop and the corona increases
\citep{2008ApJ...687L.115T,2014ApJ...787L..22A}. 
This means that the size and number of the initially induced eddies in the Mono-model would not be able to support a faster dissipation.
  
 In the multi-stranded model,
 KHI eddies extend to the whole strand region due to the mixed wave modes.
 Each individual strand is similar to the mixed driving loop in \cite{2019ApJ...870...55G}.
 However,
 since the strands are tightly packed initially,
 each deformed strand is thus strongly influenced by its neighbouring ones. 
  In this paper,
  we consider only seven strands and their radius is $0.3{\rm Mm}$.
  We mentioned that higher resolution observations showed that the fine structures in coronal loops should probably have a smaller spatial scale.
  \cite{2013Natur.493..501C} reported that the braided magnetic strands have a width of about $150{\rm km}$.
  It is true that if we fill the loop with smaller strands,
  the interactions between different strands  (see Appendix \ref{sec_appendix_B} for details) are stronger and probably lead to even more rapidly enhanced heating.
  However,
  in the scope of this paper,
  we focus on the comparison between the multi-stranded loop and its density equivalent monolithic loop,
  more strands will not influence our main conclusion that the multi-stranded loop is more efficient in starting the heating process. 
In addition,
the radius of the monolithic loop ($R=1{\rm Mm}$) is comparable to the size of the strands bundle,
although the filling factor in the multi-stranded loop is not large.
 
 The input energy flux in our models is comparable to the radiative losses of the quiet solar corona,
 though it is still smaller than $\sim100 {\rm W m^{-2}}$.
 Note that the models here are ideal and still lack some realistic solar atmosphere conditions (e.g., gravity, more realistic drivers).
 The input energy will increase when considering gravity and larger amplitude drivers (see an upcoming paper of Karampelas et al. 2019).
 Note that our models are assumed to be anchored in the lower atmosphere, 
 where the amplitude of motions should be very small \citep[$\sim5 {\rm km~s^{-1}}$ in the photosphere,][]{2010ApJ...710.1857M}.
 In our current models,
 we neglect the realistic atmosphere conditions under the corona with assuming a longitudinally uniform loop.
 If the incorporation of realistic chromosphere conditions would allow for a larger velocity amplitude
 (e.g., a non-thermal velocity of $10-20 \rm km~s^{-1}$ reported in \cite{2016ApJ...820...63B}), 
 it is indeed helpful to increase the input energy.
 However,
 we should notice that the Poynting flux not only depends on the velocity amplitude of the driver,
 but also the dynamics of the loop,
 which can influence the magnetic field perturbations at the footpoint.
 A more realistic footpoint driver according to an observed power spectrum was used in \citet{2019A&A...623A..37P} and they found that the phase mixing of Alfv{\'e}n waves is not sufficient to maintain the energy losses of the corona.
 However,
 things may change if KHI is quickly induced in a line-tied loop.
 \citet{2010ApJ...710.1857M} reported that turbulent photospheric motions can be observed by Hinode/SOT.
It is thus reasonable to consider mixed motions at the loop footpoint. 
 The realistic footpoint drivers may be complicated.
While our driver is not realistic,
 this does not influence our main conclusions aforementioned.
 We focus on the energy dissipation that depends on the configuration and the temporal evolution of loops with almost the same energy input into both models.

 In our models, 
 we solved the MHD equations in the presence of the effective numerical resistivity and viscosity  (see Appendix \ref{sec_appendix_A} for details).
 Therefore,
 the aforementioned heating is due to the numerical dissipation.
 The temperature profiles would not change even if explicitly larger resistivity and viscosity are considered \citep{2019A&A...623A..53K}.
  Even though the numerical resistivity and viscosity are significantly larger than the realistic values in the solar corona,
  we can still say that the resistive or viscous heating mechanism is effective.
  In the realistic case,
  the turbulent structures can be much smaller than those captured in the current numerical experiments,
  the effective heating can thus be achieved even with the much smaller transport coefficients.

\acknowledgments
 {The authors thank the referee for the helpful comments to improve the manuscript. This project has received funding from the European Research Council (ERC) under the European Union's Horizon 2020 research and innovation program (grant agreement No. 724326). The authors acknowledge the funding from the China Scholarship Council (CSC), the National Natural Science Foundation of China (41674172), and the GOA-2015-014 (KU Leuven). }

\clearpage
 \appendix

\section{Monolithic loop simulation with lower resolution}
\label{sec_appendix_A}

 As aforementioned,
even though we solve the nominal ideal MHD equations, 
numerical viscosity/resistivity is certainly unavoidable. 
To quantify this numerical effect, 
we have conducted a number of computations incorporating explicit viscosity and resistivity,
similar studies have been done by \citet{2017A&A...602A..74H} and \citet{2019A&A...623A..53K}. 
This is not meant to mimic the extremely small dissipation coefficients expected for the solar corona, 
but rather to offer an order-of-magnitude estimate of the numerical viscosity and resistivity.
 According to the estimate,
the dimensionless numerical resistivity and viscosity in the Ms-model and the Mono-model are of the same order ($\sim 10^{-6}$),
although the grid number that is required to resolve a single strand and the monolithic loop is different.

 To find out the influence of the numerical resistivity and viscosity,
we consider the same monolithic model with a lower resolution of $50$km (Mono-coarse model hereafter),
 such that the same number of grids can be used to resolve the Mono-coarse loop and each strand in the Ms-model.
Thus the effective numerical resistivity and viscosity are equivalent to the values in an individual strand.

\begin{figure*}
\gridline{\fig{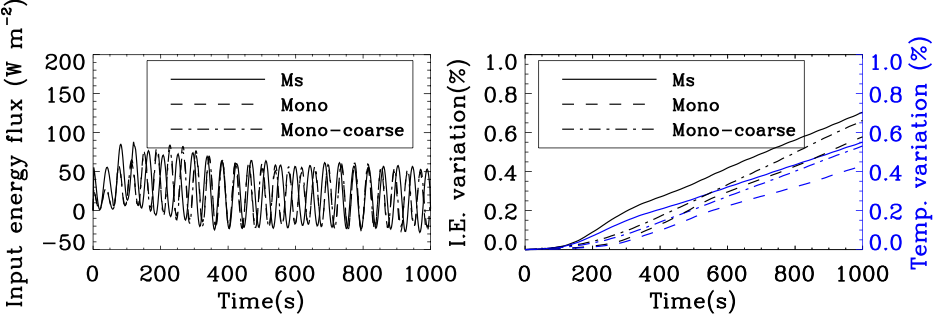}{0.9\textwidth}{}}
\caption{Left: the input energy flux variations. 
Right: percentages of volume-averaged internal energy (black) and temperature (blue) variations.}
\label{fig6}	
\end{figure*}

Similar to Figure \ref{fig2}, Figure \ref{fig6} shows the volume averaged internal energy and temperature variation with the same input energy flux in all three models.
The dissipation is slightly enhanced in the Mono-coarse model,
comparing to the Mono-model.
This means that the effective numerical resistivity/viscosity in the Mono-coarse model is larger than that in the Mono-model and thus larger than the estimated values ($\sim 10^{-6}$).
However,
both internal energy and temperature in the Mono-coarse model still show a smaller increase rate than those in the Ms-model before 300s.
This means that even if considering a more numerically dissipative monolithic loop,
the wave energy can still get a rapid dissipation in the multi-stranded loop.

 \section{Individual strand simulation}
\label{sec_appendix_B}

To reveal the interaction between different strands, 
we pick up and drive one individual strand in the Ms-model.
 We keep the same setup as in the Ms-model,
except the density distribution and the torsional driver.
The density distribution of the strand is described by
\begin{equation}
\rho_{\rm s}(r)=\rho_{\rm e}+(\rho_{\rm p}-\rho_{\rm e})\cos\left(\frac{\pi r}{2R_{\rm s}}\right),
r\le R_{\rm s},
\label{eq_strand}
\end{equation}
where $r=\sqrt{x^2+y^2}.$
We employ the same transverse driver that is described by Equation \ref{eq_v},
 but the localized torsional driver is launched in the strand region only,
\begin{equation}
v_{\theta}(r)=v_0 \sin\left[\frac{2\pi t}{P_{\rm A}(r)}\right], r\le R_{\rm s}.
\label{eq_vstrand}
\end{equation}

As aforementioned,
we expect the same energy input into different loops and thus different velocity amplitudes are employed.
For an individual strand,
the velocity amplitude is $v_0=4.3{\rm km~s^{-1}}$,
which is larger than that in the Ms-model since the torsional component is only limited in the single strand region now. 

\begin{figure*}
\gridline{\fig{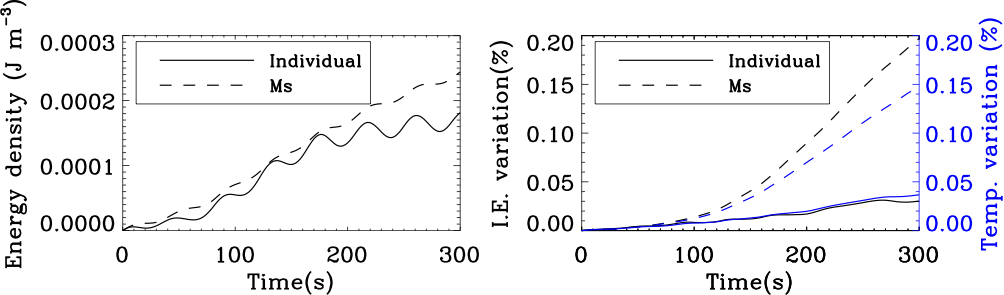}{0.9\textwidth}{}}
\caption{Left: the input energy density.
Right: percentages of volume-averaged internal energy (black) and temperature (blue) variations.  Note that the input energy density is calculated in the loop region, which is defined by $\rho(x,y,z)\geq 1.002\rho_{\rm e}$. }
\label{fig7}	
\end{figure*}

The comparison between the individual strand and the  Ms-model is shown in Figure \ref{fig7}.
Here we check the input energy density in the loop region.
The Poynting flux provided by the driver is calculated by
\begin{equation}
S(t)=-\frac{1}{V}\int^t_0\int_A \textbf{S}\cdotp d\textbf{A}dt',
\label{eq_poynting}
\end{equation}
where \textbf{S} is the Poynting flux, 
\textbf{A} is the normal surface vector of the bottom plane
and $V$ is the volume of the loop region,
which is defined by $\rho(x,y,z)\geq 1.002\rho_{\rm e}$. 
The Alfv{\'e}n component of the input energy depends on the number of the strands.
In order to obtain a more reasonable comparison between loops with different strand numbers,
the input energy is averaged in the density enhanced volume only.
We see that the input energy density in both models are almost the same before $\sim200$s in Figure \ref{fig7},
whereas the averaged internal energy and temperature have a slower increase in the individual strand model.
Therefore,
the interaction between different strands,
which is absent in the individual strand simulation,
plays an important role in dissipation.
After $\sim200$s, the input energy density in the individual strand increases slower,
due to the redistribution of the magnetic field in the bottom plane \citep[similar to the models in][]{2019ApJ...870...55G}.

\begin{figure*}
\gridline{\fig{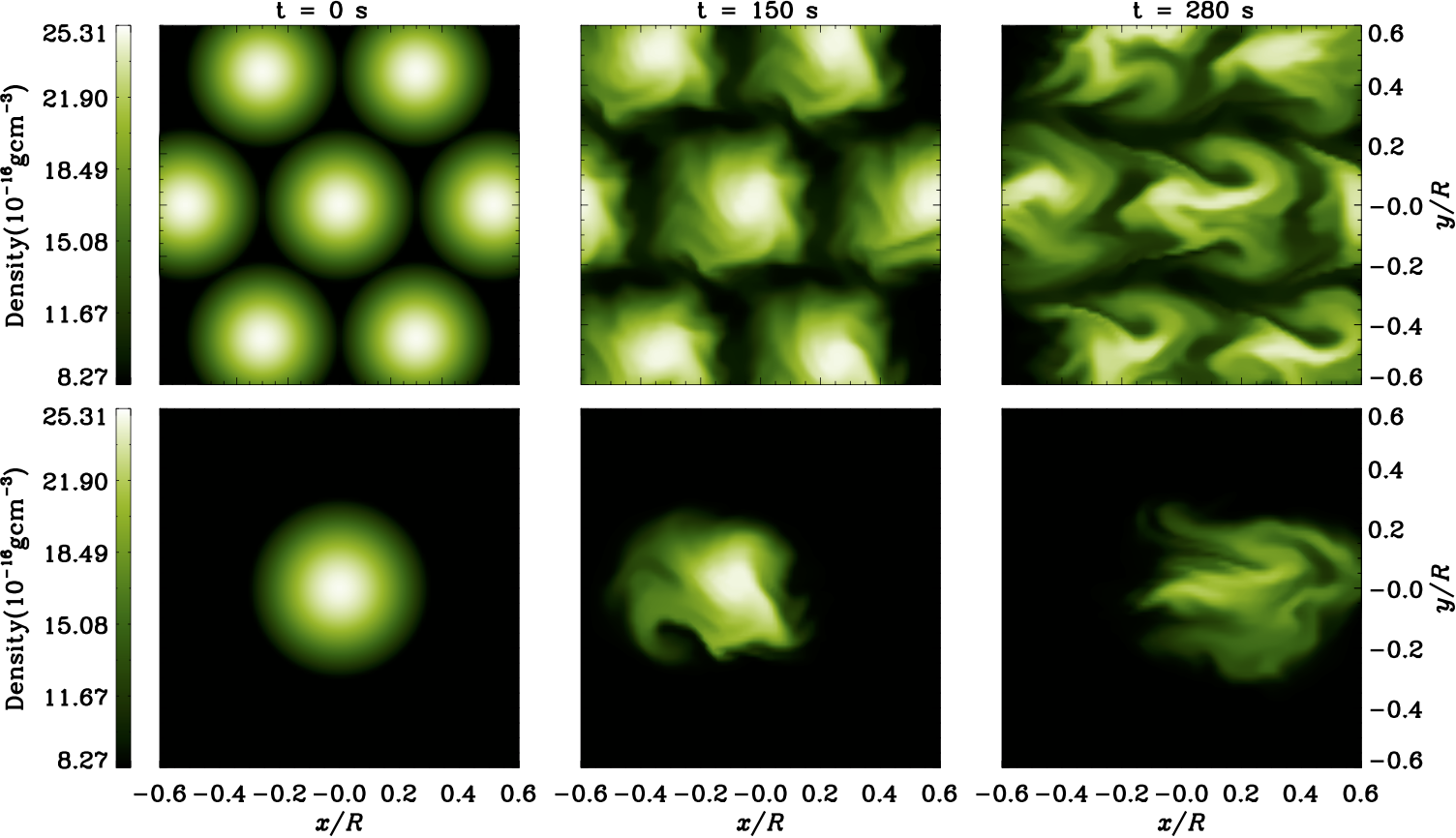}{0.9\textwidth}{}}
\caption{Snapshots of density evolution of the cross-section at loop apex for the Ms-model (upper row) and individual strand model (lower row). }
\label{fig8}	
\end{figure*}

In Figure \ref{fig8},
we see that the displacement of the central strand in the Ms-model is smaller than the individual strand.
This is due to the smaller amplitude of the driver and also the interaction with neighbouring strands in the Ms-model.
Because of the different dynamics of this strand in the Ms-model and the individual strand,
it  is not easy to compare them directly.
After several periods, for instance $t=280s$ in Figure \ref{fig8},
we can hardly recognize and pick up a single strand from the Ms-model since different strands are highly mixed.

Note that the individual strand here is not exactly a thinner ($0.3$Mm) density equivalent monolithic loop. 
The peak density of a thinner density equivalent monolithic loop should be much larger than $\rho_{\rm p}$ since the density ratio is $\alpha\approx3.42/R^2-1.16$, 
according to Equation \ref{eq_rho_mono}.
It is not straightforward to compare the individual strand to the Mono-model ($R=1$Mm) since both the peak density and the loop radius are different.
Therefore,
the influence of density contrast on the heating efficiency is unclear,
which needs a further study.

\clearpage
\bibliographystyle{apj}
\bibliography{multistranded_loop}

\begin{thebibliography}

\bibitem[Antolin et al.(2014)]{2014ApJ...787L..22A} Antolin, P., Yokoyama, T., \& Van Doorsselaere, T.\ 2014, \apjl, 787, L22 


\bibitem[Aschwanden et al.(2002)]{2002SoPh..206...99A} Aschwanden, M.~J., de Pontieu, B., Schrijver, C.~J., \& Title, A.~M.\ 2002, \solphys, 206, 99 


\bibitem[Aschwanden \& Peter(2017)]{2017ApJ...840....4A} Aschwanden, M.~J., \& Peter, H.\ 2017, \apj, 840, 4 


\bibitem[Brooks \& Warren(2016)]{2016ApJ...820...63B} Brooks, D.~H., \& Warren, H.~P.\ 2016, \apj, 820, 63 


\bibitem[Brooks et al.(2012)]{2012ApJ...755L..33B} Brooks, D.~H., Warren, H.~P., \& Ugarte-Urra, I.\ 2012, \apjl, 755, L33 


\bibitem[Cirtain et al.(2013)]{2013Natur.493..501C} Cirtain, J.~W., Golub, L., Winebarger, A.~R., et al.\ 2013, \nat, 493, 501 


\bibitem[De Moortel \& Nakariakov(2012)]{2012RSPTA.370.3193D} De Moortel, I., \& Nakariakov, V.~M.\ 2012, Philosophical Transactions of the Royal Society of London Series A, 370, 3193 


\bibitem[Edwin \& Roberts(1983)]{1983SoPh...88..179E} Edwin, P.~M., \& Roberts, B.\ 1983, \solphys, 88, 179 


\bibitem[Goossens et al.(2011)]{2011SSRv..158..289G} Goossens, M., Erd{\'e}lyi, R., \& Ruderman, M.~S.\ 2011, \ssr, 158, 289 


\bibitem[Goossens et al.(1992)]{1992SoPh..138..233G} Goossens, M., Hollweg, J.~V., \& Sakurai, T.\ 1992, \solphys, 138, 233 


\bibitem[Guarrasi et al.(2010)]{2010ApJ...719..576G} Guarrasi, M., Reale, F., \& Peres, G.\ 2010, \apj, 719, 576 


\bibitem[Guo et al.(2019)]{2019ApJ...870...55G} Guo, M., Van Doorsselaere, T., Karampelas, K., et al.\ 2019, \apj, 870, 55 


\bibitem[Heyvaerts \& Priest(1983)]{1983A&A...117..220H} Heyvaerts, J., \& Priest, E.~R.\ 1983, \aap, 117, 220 


\bibitem[Hollweg \& Yang(1988)]{1988JGR....93.5423H} Hollweg, J.~V., \& Yang, G.\ 1988, \jgr, 93, 5423 


\bibitem[Howson et al.(2017b)]{2017A&A...607A..77H} Howson, T.~A., De Moortel, I., \& Antolin, P.\ 2017, \aap, 607, A77 


\bibitem[Howson et al.(2017a)]{2017A&A...602A..74H} Howson, T.~A., De Moortel, I., \& Antolin, P.\ 2017, \aap, 602, A74 


\bibitem[Karampelas \& Van Doorsselaere(2018)]{2018A&A...610L...9K} Karampelas, K., \& Van Doorsselaere, T.\ 2018, \aap, 610, L9 


\bibitem[Karampelas et al.(2017)]{2017A&A...604A.130K} Karampelas, K., Van Doorsselaere, T., \& Antolin, P.\ 2017, \aap, 604, A130 


\bibitem[Karampelas et al.(2019)]{2019A&A...623A..53K} Karampelas, K., Van Doorsselaere, T., \& Guo, M.\ 2019, \aap, 623, A53 


\bibitem[Luna et al.(2010)]{2010ApJ...716.1371L} Luna, M., Terradas, J., Oliver, R., \& Ballester, J.~L.\ 2010, \apj, 716, 1371 


\bibitem[Luna et al.(2008)]{2008ApJ...676..717L} Luna, M., Terradas, J., Oliver, R., \& Ballester, J.~L.\ 2008, \apj, 676, 717 


\bibitem[Magyar \& Van Doorsselaere(2016a)]{2016A&A...595A..81M} Magyar, N., \& Van Doorsselaere, T.\ 2016, \aap, 595, A81 


\bibitem[Magyar \& Van Doorsselaere(2016b)]{2016ApJ...823...82M} Magyar, N., \& Van Doorsselaere, T.\ 2016, \apj, 823, 82 


\bibitem[Magyar et al.(2015)]{2015A&A...582A.117M} Magyar, N., Van Doorsselaere, T., \& Marcu, A.\ 2015, \aap, 582, A117 


\bibitem[Matsumoto \& Shibata(2010)]{2010ApJ...710.1857M} Matsumoto, T., \& Shibata, K.\ 2010, \apj, 710, 1857 


\bibitem[Mignone(2014)]{2014JCoPh.270..784M} Mignone, A.\ 2014, Journal of Computational Physics, 270, 784 


\bibitem[Mignone et al.(2007)]{2007ApJS..170..228M} Mignone, A., Bodo, G., Massaglia, S., et al.\ 2007, \apjs, 170, 228 


\bibitem[Nakariakov et al.(1999)]{1999Sci...285..862N} Nakariakov, V.~M., Ofman, L., Deluca, E.~E., Roberts, B., \& Davila, J.~M.\ 1999, Science, 285, 862 


\bibitem[Nakariakov et al.(2016)]{2016SSRv..200...75N} Nakariakov, V.~M., Pilipenko, V., Heilig, B., et al.\ 2016, \ssr, 200, 75 


\bibitem[Ofman(2009)]{2009ApJ...694..502O} Ofman, L.\ 2009, \apj, 694, 502 


\bibitem[Ofman et al.(1995)]{1995ApJ...444..471O} Ofman, L., Davila, J.~M., \& Steinolfson, R.~S.\ 1995, \apj, 444, 471 


\bibitem[Ofman \& Wang(2008)]{2008A&A...482L...9O} Ofman, L., \& Wang, T.~J.\ 2008, \aap, 482, L9 


\bibitem[Pagano \& De Moortel(2019)]{2019A&A...623A..37P} Pagano, P., \& De Moortel, I.\ 2019, \aap, 623, A37 


\bibitem[Parnell \& De Moortel(2012)]{2012RSPTA.370.3217P} Parnell, C.~E., \& De Moortel, I.\ 2012, Philosophical Transactions of the Royal Society of London Series A, 370, 3217 


\bibitem[Pascoe et al.(2011)]{2011ApJ...731...73P} Pascoe, D.~J., Wright, A.~N., \& De Moortel, I.\ 2011, \apj, 731, 73 


\bibitem[Pascoe et al.(2010)]{2010ApJ...711..990P} Pascoe, D.~J., Wright, A.~N., \& De Moortel, I.\ 2010, \apj, 711, 990 


\bibitem[Peter et al.(2013)]{2013A&A...556A.104P} Peter, H., Bingert, S., Klimchuk, J.~A., et al.\ 2013, \aap, 556, A104 


\bibitem[Poedts et al.(1990)]{1990ApJ...360..279P} Poedts, S., Goossens, M., \& Kerner, W.\ 1990, \apj, 360, 279 


\bibitem[Poedts \& Kerner(1991)]{1991PhRvL..66.2871P} Poedts, S., \& Kerner, W.\ 1991, Physical Review Letters, 66, 2871 


\bibitem[Reale et al.(2011)]{2011ApJ...736L..16R} Reale, F., Guarrasi, M., Testa, P., et al.\ 2011, \apjl, 736, L16 


\bibitem[Soler \& Terradas(2015)]{2015ApJ...803...43S} Soler, R., \& Terradas, J.\ 2015, \apj, 803, 43 


\bibitem[Taroyan \& Erd{\'e}lyi(2009)]{2009SSRv..149..229T} Taroyan, Y., \& Erd{\'e}lyi, R.\ 2009, \ssr, 149, 229 


\bibitem[Terradas et al.(2008a)]{2008ApJ...687L.115T} Terradas, J., Andries, J., Goossens, M., et al.\ 2008, \apjl, 687, L115 


\bibitem[Terradas et al.(2008b)]{2008ApJ...679.1611T} Terradas, J., Arregui, I., Oliver, R., et al.\ 2008, \apj, 679, 1611 


\bibitem[Terradas \& Ofman(2004)]{2004ApJ...610..523T} Terradas, J., \& Ofman, L.\ 2004, \apj, 610, 523 


\bibitem[Testa et al.(2002)]{2002ApJ...580.1159T} Testa, P., Peres, G., Reale, F., \& Orlando, S.\ 2002, \apj, 580, 1159 


\bibitem[Tomczyk et al.(2007)]{2007Sci...317.1192T} Tomczyk, S., McIntosh, S.~W., Keil, S.~L., et al.\ 2007, Science, 317, 1192 


\bibitem[Tripathi et al.(2009)]{2009ApJ...694.1256T} Tripathi, D., Mason, H.~E., Dwivedi, B.~N., del Zanna, G., \& Young, P.~R.\ 2009, \apj, 694, 1256 


\bibitem[Van Doorsselaere et al.(2007)]{2007A&A...471..311V} Van Doorsselaere, T., Andries, J., \& Poedts, S.\ 2007, \aap, 471, 311 


\bibitem[Van Doorsselaere et al.(2008)]{2008A&A...485..849V} Van Doorsselaere, T., Ruderman, M.~S., \& Robertson, D.\ 2008, \aap, 485, 849 


\bibitem[Withbroe \& Noyes(1977)]{1977ARA&A..15..363W} Withbroe, G.~L., \& Noyes, R.~W.\ 1977, \araa, 15, 363 




\end{thebibliography}

%%%%%%%%%%%%%%%%%%%%%%%%%%%%%%%%%%%%%%%%%%%
%%%%%%%%%%%%%%%%%%%%%%%%%%%%%%%%%%%%%%%%%%%
	
\end{document}